\documentclass[a4paper, 12pt, authoryear, times]{elsarticle} %
\biboptions{sort&compress}

\usepackage{amsmath}
\usepackage{siunitx}
\usepackage{lineno,hyperref}
\usepackage{tabularx}
\modulolinenumbers[5]

\usepackage{geometry}
\usepackage[colorinlistoftodos, textwidth=2.cm, textsize=footnotesize,
             shadow, backgroundcolor=green, linecolor=green]{todonotes}  

\makeatletter
\newcommand*{\rom}[1]{\expandafter\@slowromancap\romannumeral #1@}
\makeatother\newcommand*\diff{\mathop{}\!\mathrm{d}}
\usepackage{upgreek}
\usepackage{graphicx}
\usepackage{subfigure}
\usepackage{color}
\usepackage{caption}
\usepackage{lipsum}
\usepackage{tabularx}
\usepackage{booktabs}

\usepackage{color}

\DeclareMathAlphabet{\Ibb}{U}{msb}{m}{n}
\newcommand   {\IC}{{\ensuremath{\Ibb C}}}
\newcommand \ggp {{\ensuremath{{{\gamma}/{\gamma}^{\prime}}}}}
\newcommand \gp  {{\ensuremath{{\gamma}^{\prime}}}}
\newcommand \g   {{\ensuremath{\gamma}}}
\newcommand{\matone}{\ensuremath{\text{\textup{\textbf{I}}}}}

 \newcommand{\dV}{\text{\rm d}V}

 \newcommand{\Bb}{{\boldsymbol{\mathnormal b}}}

 \newcommand{\Bn}{{\boldsymbol{\mathnormal n}}}

 \newcommand{\BP}{{\boldsymbol{\mathnormal P}}}

 \newcommand{\Bsigma} {\ensuremath{\boldsymbol\sigma}}

 \newcommand{\Bepsilon    }{\ensuremath{\boldsymbol\epsilon}}

\newcommand{\figref}[1]{Fig.~\ref{#1}}

\begin{document}

\begin{frontmatter}

\title{Phase-field, dislocation based plasticity and damage coupled model: modelling and application to single crystal superalloys}
\author[NWPU]{Ronghai Wu\corref{corr}}
\cortext[corr]{Corresponding author.}
\ead{ronghai.wu@nwpu.edu.cn}
\author[NWPU]{Yufan Zhang}


\address[NWPU]{School of Mechanics, Civil Engineering and Architecture, Northwestern Polytechnical University, Xian, 710072, PR China}

\begin{abstract}
In the present work, we propose a novel model coupling phase-field, dislocation density based plasticity and damage. The dislocation density governing equations are constructed based on evolutions of mobile and immobile dislocations. Mechanisms including dislocation multiplication, annihilation, mobile-immobile transfer due to dislocation interactions and block of interfaces are incorporated in the model. Especially, the "swallow-gap" problem surrounding the coarsened second phase, which often appears in dislocation and phase-field coupled simulations, is solved in the present model. Moreover, the phenomenon of dislocation cutting into the second phase during tertiary creep, which has rarely been considered in previous phase-field simulations of single crystal superalloys, is successfully captured in the present model with the coupling of damage. The long range stresses induced by external loading, coherent interface misfit, plastic activity and damage, as well as the short range stresses induced by antiphase boundary, dislocation line tension and forest dislocation trapping are considered in the dynamics of the model. High temperature $\langle 001 \rangle$ creep simulations of single crystal superalloys under \SI{200}{MPa} and \SI{350}{MPa} are conducted using the coupled model and compared with experiments. The results show that simulated phase microstructures, dislocations and creep properties principally agree with experiments during the whole creep stage, in terms of both microscopic and macroscopic features.              
\end{abstract}


\end{frontmatter} 


\section{Introduction}

\begin{figure}[htp] \centering
	\includegraphics[width=0.8\columnwidth]{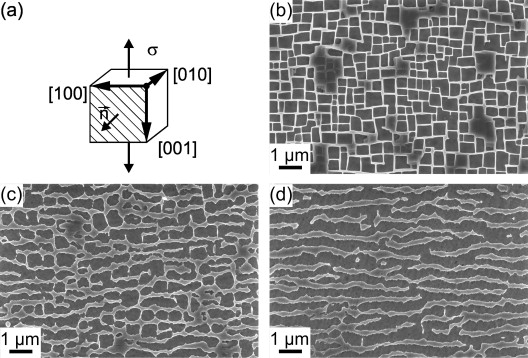}
	\caption{\label{fig:experimental microstructure} Microstructures of single crystal superalloys during $[001]$ creep at \SI{1293}{K} and \SI{160}{MPa}: (a) illustration of crystalline and loading orientations; (b) before creep; (c) after creep for \SI{19}{h}; (d) after creep for \SI{48}{h} \citep{2013_AM_Jacome}.}
\end{figure}

Single crystal Ni-based superalloys are typically used as turbine blades of advanced aircraft engines which operate at high temperatures. When the turbine spins at a constant speed, the centrifugal force exerts a constant stress which is far below the macroscopic yield stress to blades, leading to high temperature and low stress deformation (i.e. creep) of single crystal Ni-based superalloys. Understanding creep mechanisms and predicting creep properties have been a long standing demand for developing high-tech turbines. To achieve goal, macroscopic methods (e.g. macroscopic constitutive modelling) are insufficient at all, as in this case the plastic deformation takes place at external stress far below macroscopic yield stress. The microstructures must be informed. The typical microstructure evolution of single crystal Ni-based superalloys under high temperature and low stress creep is shown in \figref{fig:experimental microstructure} \citep{2013_AM_Jacome}. As can be seen, it mainly consists of cuboid $\g$ phase and channel-like $\gp$ phase surrounding the  $\g$ phase. The crystalline orientations of the $\g$ and $\gp$ phases are the same, this is why they are called "single crystal" despite that there are two phases. Typically $\langle 001 \rangle$ orientation is aligned with the centrifugal stress of blades. During creep, $\gp$ precipitates coarsen (or raft) to the direction normal to external loading, as shown in \figref{fig:experimental microstructure} (c) and (d).

To quantitatively understand the microstructure evolution and its effect on creep properties of single crystal superalloys, modelling and simulation works have been conducted using multi-scale methods such as microstructure informed constitutive modelling \citep{2021_IJP_Guo, 2019_IJP_Le, 2018_IJP_Rodas, 2018_IJP_Barba}, discrete dislocation dynamics \citep{2015_JMPS_Gao, 2017_JMPS_Gao, 2015_MSMSE_Gao}, molecular dynamics \citep{2021_MM_Yin, 2021_JAC_Wu1, 2021_JAC_Wu2} and so on. The microstructure informed constitutive models \citep{2021_IJP_Guo, 2019_IJP_Le, 2018_IJP_Rodas, 2018_IJP_Barba} are typically implemented by finite element method with the element is larger than the characteristic length of $\ggp$, indicating that $\g$ and $\gp$ phases are not explicitly resolved. The main benefit of doing so is that the simulation domain can be as big as the experimental sample and the creep properties can be predicted accurately if the model parameters are well fitted. The main drawback is that the spatial evolution of $\ggp$ phases and their interactions with dislocations cannot be investigated. In contrast, dislocation dynamics and molecular dynamics can provide solid information on dislocations and the effect of $\ggp$ on dislocations. However, the evolution of $\ggp$ cannot be simulated because of the small time scale of discrete dislocation dynamics and molecular dynamics, which restricts the investigation of effect of dislocations on $\ggp$ and the prediction of creep properties that are comparable to experiments. Phase-field method (PFM) offers a convenient framework to simulate the evolution of microstructures and predict creep properties as natural outcome of microstructure evolution (if coupling plasticity). \cite{2010_PM_Gaubert, 2011_AM_Tsukada, 2012_JMPS_Cottura} contributed the first attempts of coupling phenomenological or viscoelasticity to phase-field models to study the influence of plastic activities on the phase microstructures. Following the similar spirit, \cite{2021_CMS_Wang, 2019_AM_Wang, 2018_SM_Yang, 2017_Intermetallics_Tsukada} extended the simulations from 2D to 3D. Despite that the simulated $\ggp$ evolution agrees with experimental observations well, the fundamental carries of plasticity---dislocations---are missing in the models. Experimental observations \citep{2013_AM_Jacome, 2014_AM_Jacome, 2020_NC_Wu, 2016_AM_Wu} show that dislocations play a crucial role during the creep of single crystal superalloys. The incorporation of dislocations into phase-field model can be traced back to the work of \cite{2000_MRS_Rodney, 2001_APL_Wang, 2003_AM_Rrodney} who developed a phase-field description for discrete dislocations. That is a model at continuum scale with dislocation line still resolved, meaning that the computational cost is high and increases dramatically with the number of dislocation lines. The present authors \cite{2016_SM_Wu, 2017_IJP_Wu} contributed the first attempts on coupling phase-field and dislocation density based plastic model. \cite{2017_JAC_Wu, 2019_JAC_Wu1, 2019_JAC_Wu2} proved that the coupled models are applicable for investigating the effect of initial $\ggp$ characteristics and external loading on the evolution of phases and dislocations during creep and fatigue of single crystal superalloys. As the continuum dislocation dynamics (CDD) in those work is double checked by systematic average of discrete dislocation dynamics and functional variation of dislocation energy by \cite{2003_AM_Groma, 2015_PRL_Groma}, and numerically insured by \cite{2018_PRB_Wu, 2021_MT_Wu}, the physics of continuum dislocation dynamics is very solid. However, that PFM and CDD coupled model has two drawbacks. One is that the it currently does not have dislocation mechanisms other than dislocation glide, indicating that it can only simulate the primary creep but not the whole creep stage. Another is the "swallow-gap" problem (will be explained in detail in the following section) that most dislocation density and PFM coupled models suffer. 

Having the above mentioned in mind, we propose a model coupling PFM, dislocation density based plasticity and damage in present wok. The model is supposed to solve the "swallow-gap" problem and simulate the whole creep stage with both microscopic and macroscopic features agree with experiments. The manuscript is organized as follows. The PFM,  dislocation density based constitutive model and parameter determination will be introduced in Section \ref{sec:Modelling and parameter determination}. The stress benchmark, phase microstructure, dislocation, damage and creep property evolutions will be presented and discussed in Section \ref{sec:Results and discussions}. In the end, some conclusions are reached for the present work in Section \ref{sec:Conclusion}.                                                       

\section{Modelling and parameter determination}
\label{sec:Modelling and parameter determination}
\subsection{Phase-field model}
We use the KKS phase-field model proposed by \cite{1999_PRE_Kim, 2010_PM_Zhou} to describe the phase evolution of single crystal Ni-based superalloys. As we focus on the effect of mechanics associated mechanisms instead of chemistry associated mechanisms, a binary Ni-Al system would be proper for this purpose. The phase system is described by the composition field of Al (i.e. $c$) and order parameter field $\phi_i$, where $i=(1,2,3,4)$ as there are four different variant of the $\gp$ phase. In the $\g$ phase $c=c^\g_{\rm e}$ and $\phi_i=0$, while in the $\gp$ phase $c=c^\gp_{\rm e}$ and $\phi_i=1, \phi_j=0 (i \neq j)$, where $c^\g_{\rm e}$ and $c^\gp_{\rm e}$ are the equilibrium composition of $\g$ phase and $\gp$ phase, respectively. The $\ggp$ interface is regarded as a mix of two phases. In this way, we are informed with the phase microstructure by the $c$ and $\phi_i$ fields. The basic idea of phase-field is to construct the total free energy first, then the system evolves according to minimization of total free energy. The total free energy is formulated as a functional by
\begin{equation} 
\label{eq:total_free_energy}
F = \int_V \left ( f^{\rm chem}  + f^{\rm el} \right ) \diff V,
\end{equation}
where $f^{\rm chem}$ and $f^{\rm el}$ are the chemical energy density and elastic energy density, respectively. The formula of $f^{\rm chem}$ is given by
\begin{equation}
\label{eq: chemical_energy_density}
f^{\rm chem} = \big( 1-h(\phi_i) \big) f^{\gamma} + h(\phi_i) f^{\gamma^\prime}+ \omega g(\phi_i) + \frac{K_\phi}{2} \sum^{4}_{i=1}|\nabla\phi_i|^2, 
\end{equation}  
where $f^{\gamma}$ and $f^{\gamma^\prime}$ are the bulk free energy density of $\gamma$ phase and $\gamma^\prime$ phase, respectively. $h(\phi_i)$ and $g(\phi_i)$ are two interpolation functions. The first three terms represent the bulk energy and the last term represent the gradient energy on the right hand side of Eq. \eqref{eq: chemical_energy_density}. The formulas of $h(\phi_i)$ and $g(\phi_i)$ are given by
\begin{equation}
\label{eq: h_interpolation_function}
h(\phi_i) = \sum_{i=1}^{4} \big[\phi_i^3(6\phi_i^2 - 15\phi_i +10)\big],
\end{equation}  
\begin{equation}  
\label{eq: g_interpolation_function}
g(\phi_i) = \sum_{i=1}^{4} \phi_i^2(1-\phi_i)^2 + \theta \sum_{i,i \neq j}^{4} \phi_i^2 \phi_j^2.
\end{equation}  
The following features can be deduced: 1) $\phi_i = 0 \Rightarrow h(\phi_i)=0$ and $g(\phi_i)=0$; 2) $\phi_i = 1$ and $\phi_j = 0 \Rightarrow h(\phi_i)=1$ and $g(\phi_i)=0$; 3) $0< \phi_i < 1$ and $\phi_j =0 \Rightarrow 0<h(\phi_i)<1$ and $0 < g(\phi_i)$; 4) $0< \phi_i < 1$ and $0 <\phi_j <1 \Rightarrow 0<h(\phi_i)<1$ and $0 < g(\phi_i)$. The four features in turn represent the cases in the $\g$ phase, $\gp$ phase, $\ggp$ interphase and antiphase boundary (APB), respectively. APB is the boundary between $i$ th and $j$ th ($i \neq j$) variants of the $\gp$ phase. Note that the main difference between the third and fourth case is that the term with $\theta$ in Eq. \eqref{eq: g_interpolation_function} is zero for the former while non-zero for the later, meaning that the term with $\theta$ especially counts the addition energy caused by the APB. $f^{\gamma}$ and $f^{\gamma^\prime}$ in Eq. \eqref{eq: chemical_energy_density} are the bulk free energy density of $\gamma$ phase and $\gamma^\prime$ phase, respectively, given by   
\begin{equation}
	\label{eq:gamma_bulk_energy}
f^{\gamma}= f_0(c-c^{\rm e}_{\gamma})^2,
\end{equation}  
\begin{equation}  
	\label{eq:gamma'_bulk_energy}
f^{\gamma^\prime}= f_0(c-c^{\rm e}_{\gamma^\prime})^2,
\end{equation} 
where $f_0$ serve as energy density scale of the bulk free energy. It actually should be different for the $\gamma$ and $\gamma^\prime$ phases (i.e. $f^{\gamma}_0$ and $f^{\gamma^\prime}_0$). However, since the difference between $f^{\gamma}_0$ and $f^{\gamma^\prime}_0$ is within 10\%, we use $f_0$ for simplicity. By far, we have constructed the chemical energy density of the system, another important energy is the elastic energy density, given by
\begin{equation} 
f^{\rm el} = \frac{1}{2} \Bsigma : \Bepsilon^{\rm el},
\end{equation}
where elastic strain $\Bepsilon^{\rm el}$ and stress $\Bsigma$ follow the Hooke's law
\begin{equation} 
\label{eq: hook law}
\Bsigma = \IC : \Bepsilon^{\text{el}},
\end{equation}
where $\IC=\IC^\gp h(\phi_i) + \IC^\g (1-h(\phi_i)) $ is the stiffness tensor which is regarded as a interpolation of the stiffness tensors of the $\g$ and $\gp$ phases. In the small deformation region, the total strain can be regarded as the superposition of elastic and inelastic strains, namely 
\begin{equation} 
\Bepsilon^{\rm el} = \Bepsilon - \Bepsilon^{\rm inel}. 
\end{equation}
The inelastic strain consists of the $\ggp$ misfit strain and plastic strain, given by
\begin{equation} 
\Bepsilon^{\rm inel} = \Bepsilon^{\rm mis} + \Bepsilon^{\rm pl},
\end{equation}
\begin{equation} 
\label{eq: misfit strain}
\Bepsilon^{\rm mis}  = h(\phi_i) \bar{\epsilon}^{\rm mis} \matone,
\end{equation}
\begin{equation} 
\label{eq: plastic strain}
\Bepsilon^{\rm pl}  = \sum\limits_{k} \eta^k \boldsymbol P^{k},
\end{equation}
where $\matone$ is the identity tensor, $\eta^k$ is the plastic shear and $\BP^k$ is the projection tensor of slip system $k$ which will be introduced soon in the next subsection. Substituting Eq. \eqref{eq: hook law}-\eqref{eq: plastic strain} into the mechanical equilibrium equation $\nabla \cdot \Bsigma = \boldmath 0$ and solve it by the spectral method of micromechanics, the elastic strain, stress and energy can obtained. The evolution of Eq. \eqref{eq:total_free_energy} follows the renowned Cahn-Hilliard and Allen-Cahn equations proposed by \cite{1958_JCP_Cahn, 1979_AM_Allen}
\begin{equation} 
\label{eq: Cahn-Hilliard}
\partial_t c = \nabla \cdot \left (M_c \nabla \frac{\delta F}{\delta c} \right), 
\end{equation}
\begin{equation} 
\label{eq: Allen-Cahn}
\partial_t \phi_i = - L_\phi \frac{\delta F}{\delta \phi_i},
\end{equation}
where $M_c$ and $L_\phi$ are the mobility coefficients of diffusion and local order-disorder processes, respectively, which we assume to be isotropic and homogeneous. 

\subsection{Kinematics of dislocation density based constitutive model}
\begin{figure}[htp] \centering
	\includegraphics[width=1\columnwidth]{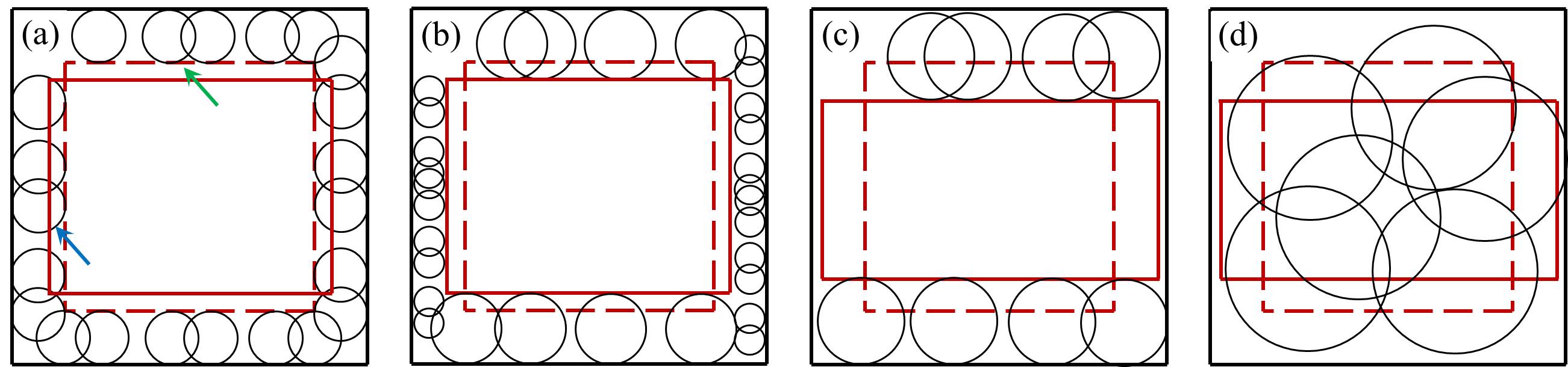}
	\caption{\label{fig:swallow-gap} Illustration of the "swallow-gap" where the initial and rafted $\ggp$ interfaces are represented by dashed and solid red lines, respectively. Dislocations are represented by loops: (a) wrong image at early creep stage; (b) correct image at early creep stage; (c) wrong image at tertiary creep stage; (d) correct image at tertiary creep stage.}
\end{figure}

Kinematics describes how dislocations and plastic shear evolve at a given velocity. The evolution rate of dislocation density can generally be described by,
\begin{equation}
\label{eq: evolution rate of dislocation density}
\partial_t \rho^k = \partial_{\eta^k} \rho^k \partial_t \eta^k,
\end{equation}
where $\rho^k$ is the total dislocation density and $\eta^k$ is the plastic shear of slip system $k$. There are two different ideas to develop dislocation-based constitutive modes in the framework of continuum mechanics. One is to directly derive $\partial_t \rho^k$ from energy functional derivative of dislocation system or systematic coarse-graining of discrete dislocation dynamics \citep{2003_AM_Groma, 2015_PRL_Groma, 2007_PM_Hochrainera}. Another is to first develop the formulations of $\partial_{\eta^k} \rho^k$ and $\partial_t \eta^k$, then the formulation of $\partial_t \rho^k$ is fixed consequentially. The former leads to the so called "continuum dislocation dynamics" describing physical dislocation density motion, which is robust from physics point of view but expensive from numerics point of view. The later leads to dislocation density evolution which is more phenomenological but computationally faster. Taking the later idea and following the spirit of \cite{1981_AM_Mecking, 1990_AM_Kubin}, together with the consideration of experimental observations of single crystal Ni-based superalloys, we propose the formulas of $\partial_{t} \rho^k$ as
\begin{equation}
\label{eq: evolution rate of mobile dislocation density extended}
\partial_t \rho^k_\text{m} = \partial_\eta \rho^k_\text{m} \partial_t \eta^k = \left[ -C_2 \rho^k_\text{m} \rho^k_\text{im} - \frac{C_3 }{b^k d_\gamma}\rho^k_\text{m} +  \frac{\xi^k C_1}{(d_\gamma^k)^4} \left( 1 - \frac{\rho_m}{\rho_\text{ref}} \right ) \right] v^k_\text{g} b^k \text{sign}(\tau^k_\text{d}),
\end{equation}  
\begin{equation}
\label{eq: evolution rate of immobile dislocation density extended}
\partial_t \rho^k_\text{im} = \partial_\eta \rho^k_\text{im} \partial_t \eta^k = \left[ C_2 \rho^k_\text{m} \rho^k_\text{im} + \frac{C_3}{b^k  d_\gamma}\rho^k_\text{m} - C_4 (\rho^k_{im})^2 \right] v^k_\text{g} b^k \text{sign}(\tau^k_\text{d}),
\end{equation}   
where $\rho^k_\text{m}$ is the density of mobile dislocations, $\rho^k_\text{im}$ is the density of immobile dislocations, $b$ is the magnitude of Burgers vector and $d_\gamma$ is the average width of $\g$ channels. $C_1$, $C_2$, $C_3$ and $C_4$ are four nondimentional parameters with positive values. The sum of $\rho^k_\text{m}$ and $\rho^k_\text{im}$ gives the total dislocation density of slip system $k$ by $\rho^k=\rho^k_\text{m}+\rho^k_\text{im}$. $\rho=\sum_k \rho^k$ gives the total dislocations density of the whole domain and $\langle \rho \rangle =\frac{1}{V}\int_v \rho \diff v$ gives the averaged or macroscopic total dislocations density of the whole domain, where $v$ is the volume of numeric grid and $V$ is the volume of the whole domain. $v^k_\text{g}$ is the dislocation velocity. $b^k$ is the Burgers vector. $\xi^k$ is a indicator which equals one if $\text{sign}(\tau^k_\text{d})>0$ and zero otherwise. Obviously, Eq. \eqref{eq: evolution rate of mobile dislocation density extended} and  Eq. \eqref{eq: evolution rate of immobile dislocation density extended} contain the Orowan equation which links the plastic shear, mobile dislocation density and velocity by
\begin{equation} 
\label{eq: Orowan shear rate final}
\partial_t \eta^k = \rho^k_\text{m} v^k_\text{g} b^k.
\end{equation}
Let us first understand the physics behind Eq. \eqref{eq: evolution rate of mobile dislocation density extended} and Eq. \eqref{eq: evolution rate of immobile dislocation density extended}. The first term on the right hand side of Eq. \eqref{eq: evolution rate of mobile dislocation density extended}, $- C_2 \rho^k_\text{m} \rho^k_\text{im}$, represents the reduction of mobile dislocation density due to mechanisms such as trapping or capturing of mobile dislocations by immobile dislocations. These mobile dislocations become immobile dislocations and hence a corresponding production term of immobile dislocations, $C_2 \rho^k_\text{m} \rho^k_\text{im}$, appears on the right hand side of Eq. \eqref{eq: evolution rate of immobile dislocation density extended}. The second term on the right hand side of Eq. \eqref{eq: evolution rate of mobile dislocation density extended} and Eq. \eqref{eq: evolution rate of immobile dislocation density extended} stands for the transition of mobile dislocations to immobile dislocations due to the block of $\ggp$ interfaces. Note that according to the formulation of the second term, the narrower the $\g$ channel, the sooner the mobile-to-immobile transition take place, which also agrees with experimental observations. The third term on the right hand side of Eq. \eqref{eq: evolution rate of mobile dislocation density extended} solve the "swallow-gap" problem surrounding the rafted $\gp$ precipitate during simulation, as illustrated in \figref{fig:swallow-gap} (a) and (b). When the $\gp$ precipitate rafts from the dashed red line morphology to the solid red line morphology, dislocations should move with $\ggp$ interfaces according to experiments (see \figref{fig:swallow-gap} (b)), not stay stationary and be swallowed by the vertical $\ggp$ interfaces (see the blue arrow in \figref{fig:swallow-gap} (a)) and leaving a dislocation-free gap near the horizontal $\ggp$ interfaces (see the green arrow in \figref{fig:swallow-gap} (a)). However, the "swallow-gap" phenomenon illustrated by \figref{fig:swallow-gap} (a), especially the swallowing of dislocations at vertical $\ggp$ interfaces which is not allowed by anti-phase boundary energy before tertiary creep stage, has became a common problem in the dislocation density based phase-field simulation of superalloys \cite{2020_PNSMI_Wu,2017_IJP_Wu,2019_JAC_Wu1}. The third term on the right hand side of Eq. \eqref{eq: evolution rate of mobile dislocation density extended} solves this "swallow-gap" problem in the following way: 1) in the $\g$ phase $\tau^k_\text{APB}$ (see \eqref{eq:APB stress}) is zero and $\xi^k$ is one, the third term brings multiplication of dislocations at the horizontal gap, but the third term decreases as the mobile dislocation density increases because of $\frac{\rho_\text{m}}{\rho_\text{ref}}$ and eventually balances the first two terms; 2) in the $\gp$ phase $\tau^k_\text{APB}$ is not zero, $\xi^k$ is zero if $\tau^k_\text{APB}>|\tau^k|$ and $\tau^k_\text{APB}$ transfers mobile dislocations swallowed at the horizontal $\ggp$ interfaces to immobile dislocations by the first two terms on the right side of Eq. \eqref{eq: evolution rate of mobile dislocation density extended}, then the third term on the right side of \eqref{eq: evolution rate of immobile dislocation density extended} further reduces the swallowed dislocations. According to experimental observation, the tertiary creep is accomplished by dislocations cutting in the $\ggp$, as illustrated in \figref{fig:swallow-gap} (c) and (d). This is also ensured by the third term on the right hand side of Eq. \eqref{eq: evolution rate of mobile dislocation density extended}: $\xi^k$ is one if $\tau^k_\text{APB}<|\tau^k|$ and the big $\tau^k$ leads to the multiplication of dislocations inside the $\gp$ phase.

\subsection{Dynamics of dislocation density based constitutive model}
Dynamics describes how different stresses affect the velocity of dislocation evolution. The dynamic equations should recover some key features observed in experiments. In the present work we propose the stress formula as
\begin{equation}
\label{eq: effective stress}
\tau_\text{eff}^k = \left
\{ 
\begin{array}{ll} \;| \tau^{k}_\text{d} | - \tau^{k}_\text{f} - \tau^{k}_\text{l}  &\text{if}\;\; | \tau^{k}_\text{d}| > \tau^{k}_\text{f} + \tau^{k}_\text{l}  \\
0 & \textrm{else,}
\end{array}
\right.
\end{equation}
where $\tau^{k}_\text{d}$ is the diving stress, $\tau^{k}_\text{f}$ is the friction stress, $\tau^{k}_\text{l}$ is the line tension stress. The velocity $v_\text{g}^k$ is proposed as
\begin{equation}
v_\text{g}^k = v_0 \left (\frac{\tau_\text{eff}^k}{\tau_\text{ref}} \right)^m \text{sign}(\tau^k_\text{d})
\end{equation}
where the $v_0$ and $\tau_\text{ref}$ are introduced for dimensional purpose. In the following, the origin of different stress terms in Eq. \eqref{eq: effective stress} will be explained in detail. The formula of $\tau^{k}_\text{d}$ is given by
\begin{equation}
\label{eq: driving stress}
\tau^{k}_\text{d} = (|\tau^k|-\tau^{k}_\text{APB}) \sum_i \phi_i + (1-\sum_i \phi_i)\tau^k ,
\end{equation}
where $\tau^k$ is the resolved shear stress and $\tau^{k}_\text{APB}$ is the APB stress. \eqref{eq: driving stress} has the feature that in the $\g$ phase $\tau^{k}_\text{d} = \tau^k$ and in the $\gp$ phase $\tau^{k}_\text{d} = |\tau^k|-\tau^{k}_\text{APB}$. $\tau^k_\text{APB}$ comes from the fact that an $i$ th variant converts to $j$ th ($j\neq i$) variant of the $\gp$ phase when dislocations in the $\g$ phase shear into the $\gp$ phase, and the sheared plane is actually an boundary between the $i$ th and $j$ th ($j\neq i$) variants (i.e. APB) with much bigger energy than the $\ggp$ interface, which serves a resistance stress preventing dislocations in the $\g$ phase to cut into the $\gp$ phase. Before $\tau^k$ is big enough to let dislocations cut through $\gp$ phase, dislocations swallowed by the vertical $\ggp$ interface during rafting are driven out by the $\tau^k_\text{APB}$. This is why $\tau^k_\text{APB}$ is regard as a part of the driving stress for dislocations inside the $\ggp$ phase in Eq. \eqref{eq: driving stress}, as has been explained in detail with the assistance of \figref{fig:swallow-gap}. 
 The formula of $\tau^{k}_\text{APB}$ is given by
\begin{equation}
\label{eq:APB stress}
\tau^k_\text{APB} = \frac{\cal E_\text{APB}}{b^k} \sum_i \phi_i,
\end{equation} 
where $\cal E_\text{APB}$ is the APB energy and $\sum_i \phi_i$ ensures that the APB stress exists in the $\gp$ phase only. The formula of $\tau^k$ is given by
\begin{equation}
\label{eq:resolved stress}
\tau^k=\frac{1}{1-D}\Bsigma:\BP^k,
\end{equation} 
where $\frac{1}{1-D}$ represents the increment of resolved stress due to the reduction of contact area by voids and cracks. $D$ initially equals $0$ before deformation and increases as the deformation goes on, reaching $1$ eventually at rapture. $\frac{1}{1-D}$ is the main reason that the APB resistance Eq. \eqref{eq:APB stress} cannot be overcame at the early creep stage but can be overcame at the late creep stage. Since the local elastic energy is often used as criteria for crack and void formation, the evolution of $D^k$ is set to be      
\begin{equation}
\label{eq:damage evolution}
\partial_t D = D_0 \Bsigma:\Bepsilon^\text{el}, 
\end{equation} 
where $D_0$ is a coefficient controlling the kinetic of damage evolution. The formula of $\tau^k_\text{l}$ is proposed as 
\begin{equation}
\label{eq:line tension stress}
\tau^k_\text{l} = \frac{\rho}{\rho_\text{ref}} \frac{G^k b^k}{0.5d_\g} \left( 1-\sum_i\phi_i \right ) 
\end{equation} 
where $G^k$ is the shear modulus. $\tau^k_\text{l}$ comes from that fact that curved dislocations need to overcome the maximum line tension stress to multiply in the $\g$ channels. Since $1-\sum_i\phi_i=0$ in the $\gp$ phase and $1-\sum_i\phi_i=1$ in the $\g$ phase, $1-\sum_i\phi_i$ serves as a spatial indicator which indicates that the maximum line tension stress is considered only in the $\g$ phase. Because when dislocations cut into the $\gp$ phase, the dislocation line curvature increases dramatically, meaning that the line tension stress decreases to negligible value. $\frac{\rho}{\rho_\text{ref}}$ comes from the fact as more and more dislocations are blocked and pile up at the $\ggp$ interface, the bowing out space decreases and resistance of dislocation bowing out increases. It is worth mentioning that \cite{2016_JMPS_Cottura, 2017_IJP_Wu} proposed back stress terms in phase-field and dislocation-based plasticity coupled models for single crystal Ni-based superalloys. \cite{2003_PM_Forest} has shown that the back stress in continuum models of straight dislocations actually mimics the hardening by line tension stress of curved dislocations in the case that dislocation motion is confined in narrow channels. Hence, the physics behind the line tension stress in the present and the back stress proposed by \cite{2016_JMPS_Cottura, 2017_IJP_Wu} is the same. The formula of is given by
\begin{equation}
\label{eq:Tarlor friction stress simplified}
\tau^k_\text{f}= \alpha G^k b^k \sqrt{\rho}.
\end{equation}  
where $\alpha$ is a value typically in the range of $0.2-0.4$. The Taylor friction stress represents the mutual trapping of positive and negative dislocations into dipolar or multipolar configurations. 

\subsection{Parameter identification and simulation setup}
We do creep loading simulations at \SI{1253}{K}, so all the parameters have their values at \SI{1253}{K}. The equilibrium composition of the $\g$ and $\gp$ can be obtain from phase diagram. Other parameters in Eq. \eqref{eq: chemical_energy_density}-\eqref{eq:gamma'_bulk_energy} are mutually restricted and determined by 1D numeric experiments with rules \citep{2008_PHD_Zhou, 2016_JMPS_Cottura}: 1) there are more than 5 grids in the $\ggp$ interface to avoid numeric pinning; 2) the $\ggp$ interface energy is in the range of $10 \sim 100$ \SI{}{J m^{-2}} with the bulk energy and gradient energy contribute equally; 3) the APB energy is more than two times of the $\ggp$ interface energy. The stiffness tensor and misfit of the $\ggp$ interface in Eq. \eqref{eq: hook law}-\eqref{eq: misfit strain} can be found in experimental or lower scale calculation literatures. The main slip systems are $\{011\}\langle 11\bar{1} \rangle$ for FCC crystals, as four of them have zero resolved shear stress under $\langle 001 \rangle$ loading, we only take the eight active slip systems whose projection tensors are listed in Table \ref{tab:Fatigue_Simulation_Parameters}. The mobility coefficient $M_c$ in Ep. \eqref{eq: Cahn-Hilliard} is determined such that the Ni-Al inter-diffusion is recovered, because Ep. \eqref{eq: Cahn-Hilliard} eventually abides by the Fick's laws of diffusion. The mobility coefficient $L_\phi$ in Ep. \eqref{eq: Allen-Cahn} takes a much larger value than $M_c$ by $L_\phi \gg M_c/\Delta l$, where $\Delta l$ is the grid spacing. Because from physics point of view, local order-disorder is much fast than diffusion. The parameters associated with dislocation density based constitutive modes in Eq. \eqref{eq: evolution rate of mobile dislocation density extended}-\eqref{eq:Tarlor friction stress simplified} are mutually constrained. The difference between any of the two terms (i.e. the terms with $C_1 \sim C_4$) on the right hand side of $\partial_{\eta^k} \rho^k_{\rm m}$ and $\partial_{\eta^k} \rho^k_{\rm im}$ cannot always exceed two orders of magnitude, otherwise the term which is constantly two orders of magnitude smaller than the rest can be abandoned from the formulas. The $v_0$, $m$ and $\tau_{\rm ref}$ in the dislocation velocity formula, which have influence on the shape of $\eta$ and hence $\langle \Bepsilon^{\rm pl} \rangle$, can be adjusted with the experimental data of $\langle \Bepsilon^{\rm pl} \rangle$. The $\rho_{\rm ref}$ has influence on the shape of $\langle \rho \rangle$ which is typically around \SI{1e13}{m^{-2}} initially and increases to $3\times10^{13} \sim$ \SI{3e14}{m^{-2}} after primary creep according to experiments.  
      
Having the above mentioned strategies in mind, the parameters are determined as follows: $K_\phi = \SI{20e-10}{J m^{-1}}$, $\theta = \SI{10}{}$, $\omega = \SI{3.9e6}{J m^{-3}}$, $M_c= \SI{1.e-26}{J^{-1} m^{5} s^{-1}}$, $L_\phi = \SI{5e-9}{J^{-1} m^{3} s^{-1}}$, $C_1 = \SI{9.6}{}$, $C_2 = \SI{1020}{}$, $C_3 = \SI{0.072}{}$, $C_4 = \SI{2400}{}$, $\rho_{\rm ref} = \SI{3e14}{m^{-2}}$, $v_0 = \SI{4e-10}{m s^{-1}}$, $m=0.5$, $\tau_{\rm ref} = \SI{0.01}{GPa}$, $b = \SI{0.25e-9}{m}$, $d_\g = \SI{e-7}{m}$, $D_0 = \SI{3.5e-3}{GPa^{-1} s^{-1}}$, $\alpha = \SI{0.3}{}$, $\cal E_{\rm APB} = \SI{0.1}{J m^{-2}}$, $\IC^\g_{11} = \SI{210}{GPa}$, $\IC^\g_{12} = \SI{154}{GPa}$, $\IC^\g_{44} = \SI{84}{GPa}$, $\IC^\gp_{11} = \SI{220}{GPa}$, $\IC^\gp_{12} = \SI{140}{GPa}$, $\IC^\gp_{44} = \SI{122}{GPa}$, $\bar{\epsilon}^{\rm mis} = -0.003$. The governing equations are solved by the spectral method \citep{2001_AM_Hu}.  

\begin{table}
	\centering \setlength\tabcolsep{15pt}%
	\captionof{table}{Projection tensors of eight active slip systems under $[010]$ loading} 
	\label{tab:Fatigue_Simulation_Parameters}
	\begin{tabular}{c c c}
		\hline
		$k$  & $\Bn$ and $\Bb$ & $\BP$  \\
		\hline
		\specialrule{0em}{3pt}{3pt}
		1  & 
		$\Bn=\frac{1}{\sqrt{3}}\begin{bmatrix}
		1 & 1 & -1
		\end{bmatrix}$ 
		,
		$\Bb=\frac{b}{\sqrt{2}}\begin{bmatrix}
		0 & 1 & 1
		\end{bmatrix}$  
		& 
		$\frac{1}{\sqrt{6}}\begin{bmatrix}
		0 & 0.5 & 0.5\\
		0.5 & 1 & 0\\
		0.5 & 0 & -1
		\end{bmatrix}$    \\
		\specialrule{0em}{3pt}{3pt}
		
		2  & 
		$\Bn=\frac{1}{\sqrt{3}}\begin{bmatrix}
		-1 & 1 & -1
		\end{bmatrix}$ 
		,
		$\Bb=\frac{b}{\sqrt{2}}\begin{bmatrix}
		1 & 1 & 0
		\end{bmatrix}$  
		& 
		$\frac{1}{\sqrt{6}}\begin{bmatrix}
		-1 & 0 & -0.5\\
		0 & 1 & -0.5\\
		-0.5 & -0.5 & 0
		\end{bmatrix}$    \\
		\specialrule{0em}{3pt}{3pt}
		
		3  & 
		$\Bn=\frac{1}{\sqrt{3}}\begin{bmatrix}
		-1 & 1 & 1
		\end{bmatrix}$ 
		,
		$\Bb=\frac{b}{\sqrt{2}}\begin{bmatrix}
		0 & 1 & -1
		\end{bmatrix}$  
		& 
		$\frac{1}{\sqrt{6}}\begin{bmatrix}
		0 & -0.5 & 0.5\\
		-0.5 & 1 & 0\\
		0.5 & 0 & -1
		\end{bmatrix}$    \\
		\specialrule{0em}{3pt}{3pt}
		
		4  & 
		$\Bn=\frac{1}{\sqrt{3}}\begin{bmatrix}
		1 & 1 & 1
		\end{bmatrix}$ 
		,
		$\Bb=\frac{b}{\sqrt{2}}\begin{bmatrix}
		-1 & 1 & 0
		\end{bmatrix}$  
		& 
		$\frac{1}{\sqrt{6}}\begin{bmatrix}
		-1 & 0 & -0.5\\
		0 & 1 & 0.5\\
		-0.5 & 0.5 & 0
		\end{bmatrix}$    \\
		\specialrule{0em}{3pt}{3pt}
		
		5  & 
		$\Bn=\frac{1}{\sqrt{3}}\begin{bmatrix}
		1 & 1 & 1
		\end{bmatrix}$ 
		,
		$\Bb=\frac{b}{\sqrt{2}}\begin{bmatrix}
		0 & 1 & -1
		\end{bmatrix}$  
		& 
		$\frac{1}{\sqrt{6}}\begin{bmatrix}
		0 & 0.5 & -0.5\\
		0.5 & 1 & 0\\
		-0.5 & 0 & -1
		\end{bmatrix}$    \\
		\specialrule{0em}{3pt}{3pt}	
		
		6  & 
		$\Bn=\frac{1}{\sqrt{3}}\begin{bmatrix}
		-1 & 1 & 1
		\end{bmatrix}$ 
		,
		$\Bb=\frac{b}{\sqrt{2}}\begin{bmatrix}
		1 & 1 & 0
		\end{bmatrix}$  
		& 
		$\frac{1}{\sqrt{6}}\begin{bmatrix}
		-1 & 0 & 0.5\\
		0 & 1 & 0.5\\
		0.5 & 0.5 & 0
		\end{bmatrix}$    \\
		\specialrule{0em}{3pt}{3pt}
		
		7  & 
		$\Bn=\frac{1}{\sqrt{3}}\begin{bmatrix}
		-1 & 1 & -1
		\end{bmatrix}$ 
		,
		$\Bb=\frac{b}{\sqrt{2}}\begin{bmatrix}
		0 & 1 & 1
		\end{bmatrix}$  
		& 
		$\frac{1}{\sqrt{6}}\begin{bmatrix}
		0 & -0.5 & -0.5\\
		-0.5 & 1 & 0\\
		-0.5 & 0 & -1
		\end{bmatrix}$    \\
		\specialrule{0em}{3pt}{3pt}
		
		8  & 
		$\Bn=\frac{1}{\sqrt{3}}\begin{bmatrix}
		1 & 1 & -1
		\end{bmatrix}$ 
		,
		$\Bb=\frac{b}{\sqrt{2}}\begin{bmatrix}
		-1 & 1 & 0
		\end{bmatrix}$  
		& 
		$\frac{1}{\sqrt{6}}\begin{bmatrix}
		-1 & 0 & 0.5\\
		0 & 1 & -0.5\\
		0.5 & -0.5 & 0
		\end{bmatrix}$    \\
		\specialrule{0em}{3pt}{3pt}
		\hline
	\end{tabular}
\end{table}

\section{Results ans discussions}
\label{sec:Results and discussions}
\subsection{Stress benchmark}
\begin{figure}[htp] \centering
	\includegraphics[width=\columnwidth]{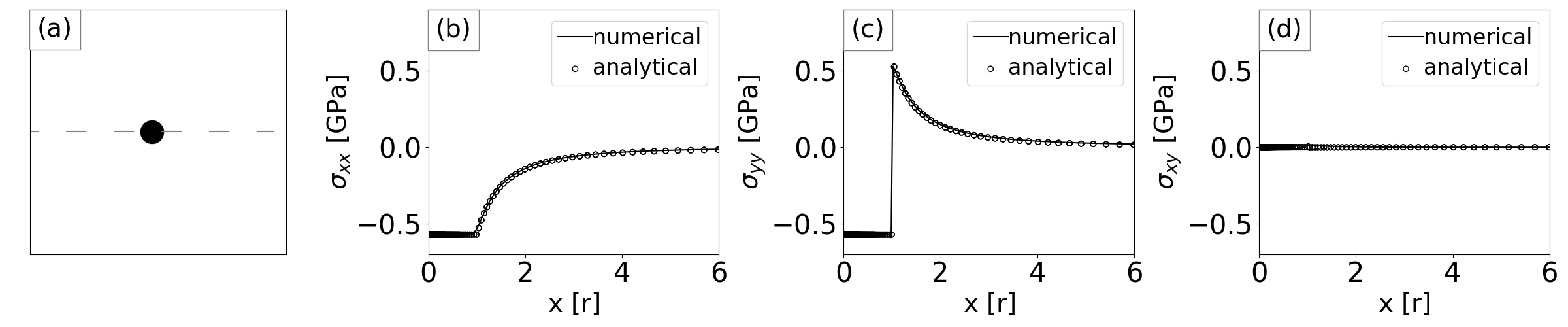}
	\caption{\label{fig: inclusion benchmark} Benchmark of stress induced by inclusion: (a) 2D setup; (b)-(d) comparison between numerical and analytical results cutting along the dashed line, where: (b) $\sigma_{xx}$, (c) $\sigma_{yy}$, (d) $\sigma_{xy}$, and r is the radius of inclusion.}
\end{figure}

\begin{figure}[htp] \centering
	\includegraphics[width=\columnwidth]{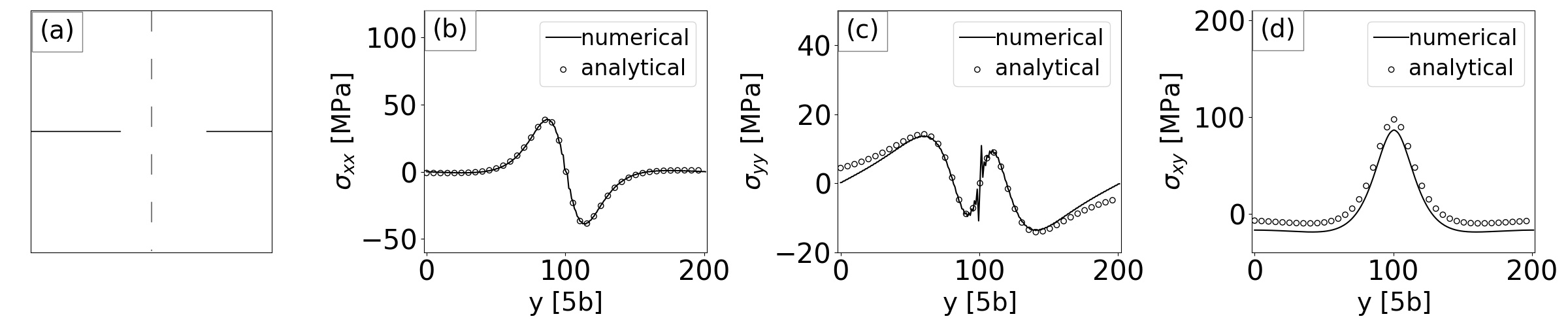}
	\caption{\label{fig: dislocation benchmark} Benchmark of stress induced by inclusion: (a) 2D setup; (b)-(d) comparison between numerical and analytical results cutting along the dashed line, where: (b) $\sigma_{xx}$, (c) $\sigma_{yy}$, (d) $\sigma_{xy}$.}
\end{figure}

As stress and strain are crucial issues for the present work, we fist validate the numerical stresses by analytical solutions, as shown in \figref{fig: inclusion benchmark} and \figref{fig: dislocation benchmark}. Both cases are set to be isotropic and homogenous, because the analytical solutions of them have been widely accepted \citep{2007_AM_Gururajan, 2012_CG_Meng, 2017_book_Anderson}. \figref{fig: inclusion benchmark} presents the case that a small circular inclusion is located at the center of the domain (see \figref{fig: inclusion benchmark}(a)). The diagonal components of the misfit strain between inclusion and matrix are set be $0.01$ and the off-diagonal components are set to be zero, such that the inclusion has similar misfit strain feature with the $\ggp$. The elastic constants are set to be $\IC_{11} = \SI{140}{GPa}$,  $\IC_{12} = \SI{60}{GPa}$ and  $\IC_{44} = \SI{40}{GPa}$ to make material isotropic. The three stress components are plotted along $y=0.5L_y$ (i.e. the dashed line) of the domain (see \figref{fig: inclusion benchmark}(b)-(d)), where $L_y$ is the vertical length of the domain. \figref{fig: dislocation benchmark} presents the case that plastic shear of one Burger vector is located at the horizontal center of the domain. One can understand the situation by imaging a positive dislocation glides from the left boundary and parks at the edge of plastic shear, while a negative dislocation glides from the right boundary and parks at the edge of plastic shear, therefore generating plastic shear as illustrated in \figref{fig: dislocation benchmark}(a). In this case the diagonal components of the inelastic strain are zero and the off-diagonal components are $0.5b/\Delta l$, where $\Delta l$ is the grid spacing which we set to be $5b$. The three stress components are plotted along $x=0.5L_x$ (i.e. dashed line) of the domain (see \figref{fig: dislocation benchmark}(b)-(d)), where $L_x$ is the horizontal length of the domain. As can be seen, the numerical stresses in both cases agree with analytical solutions very well.        

\subsection{Phase microstructure evolution}
\begin{figure}[htp] \centering
\centering
\includegraphics[width=1\columnwidth]{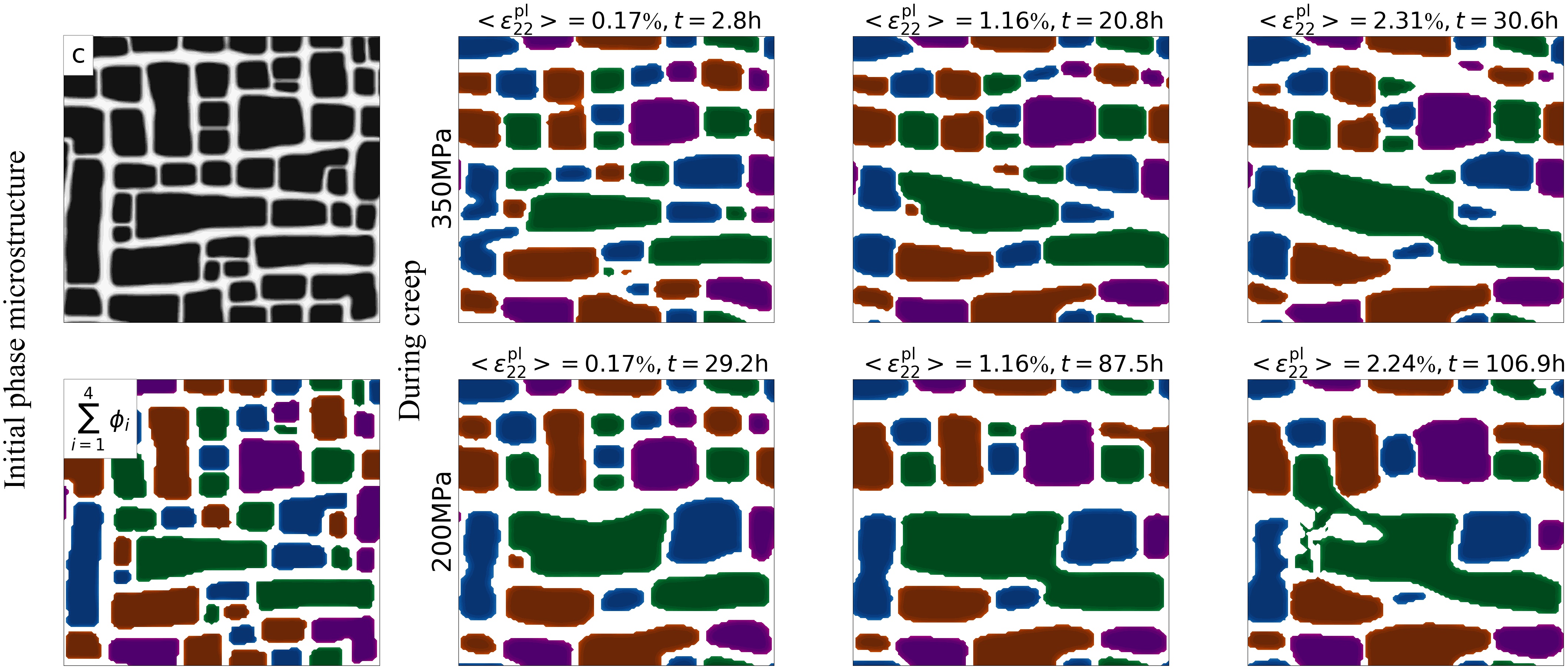}
\caption{\label{fig: phase microstructure} Phase microstructures before and during creep.}
 \end{figure}

The phase microstructures before and during creep are shown in \figref{fig: phase microstructure}. The four $\gp$ variants are distinguished by four different colors. The initial phase microstructure for creep is generated by setting composition field around $c=0.209$ with small fluctuation which leads to spinodal decomposition type of precipitating of the $\gp$ phase from supersaturation $\g$ phase, according to the chemical free energy of phase microstructure. As the precipitation goes on, the well-formed $\ggp$ phase microstructure is obtained as the initial microstructure for creep. The simulated well-formed phase microstructure agrees with the experimental phase microstructure. Specifically, the representative shapes of the $\gp$ precipitates including square, rectangle and L-shape observed in experiments \citep{2015_MMTA_Fleischmann, 2015_MSMSE_Gao} also present in the simulated phase microstructure. \SI{200}{MPa} and \SI{350}{MPa} tensile creep loadings are applied along the vertical direction (i.e. $\langle 001 \rangle$). Under \SI{350}{MPa} loading, it is obvious that the $\ggp$ precipitates raft to the horizontal direction. A notable feature during early creep (i.e. $\langle \epsilon^{\rm pl}_{22} \rangle=0.17\%$) is that the vertically elongated $\ggp$ precipitates are splitting. As the creep goes on and reaches the transition between secondary and tertiary creep ($\langle \epsilon^{\rm pl}_{22} \rangle=1.16\%$), the horizontal $\ggp$ interfaces become increasingly wavy, but the $\gp$ precipitates at different layers are not connected yet. However, during the tertiary creep ($\langle \epsilon^{\rm pl}_{22} \rangle=2.31\%$), it is notable that $\gp$ precipitates at different layers try to connect each other, forming a microstructure that the $\g$ channels are surrounded by $\gp$ precipitates. Such a phenomenon is the renowned topological inversion during tertiary creep of single crystal superalloys, because initially the phase microstructure is that the $\gp$ precipitates surrounded by $\g$ channels but now it is the other way round. Under \SI{200}{MPa} loading, the principal features of phase microstructure evolution are similar with those under \SI{200}{MPa} loading. The main differences are: 1) it take much longer time to reach the same macroscopic plastic strain in the case of \SI{200}{MPa} loading; 2) the rafting of $\gp$ precipitates is less severe in the case of \SI{200}{MPa} loading, due to the lower mechanical energy. The phase microstructure evolution in \figref{fig: phase microstructure} agree with massive experimental observations of single crystal superalloys under similar loading conditions \citep{2011_PM_Viguier, 2006_MSEA_Shui, 2012_conference_Fedelich, 2000_conference_Miura}. The mechanical energy induced phase microstructure evolution can be explained based on the dislocation and damage evolutions as follows.               

\subsection{Dislocation and damage evolutions}
\begin{figure}[htp] \centering
	\includegraphics[width=\columnwidth]{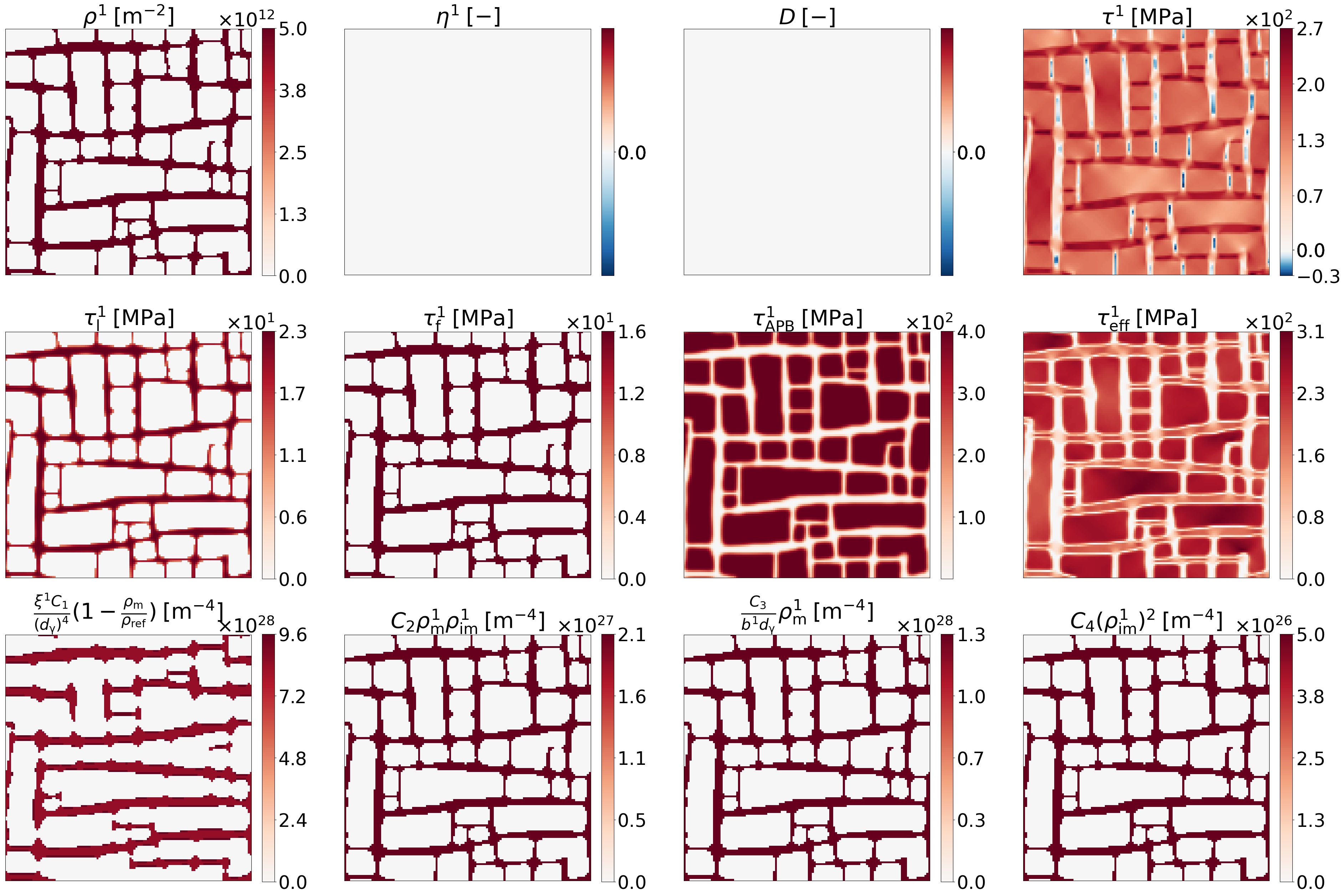}
	\caption{\label{fig: 350MPa_t=0h} Dislocation of slip system one and damage associated fields at $\langle \epsilon^{\rm pl}_{22} \rangle=\SI{0}{\%}$ and $t=\SI{0}{h}$ under \SI{350}{MPa}.}
\end{figure}

The initial dislocation of slip system one and damage associated fields are shown in \figref{fig: 350MPa_t=0h}. We always use red color and blue color for positive and negative values, respectively, and the white color represents zero. The maximum and minimum values of each color bar are the real maximum and minimum values of the corresponding field. The initial dislocation density of slip system one at the $\g$ channels is about \SI{5e12}{m^{-2}}, leading to average dislocation density $\langle \sum_k \rho^k \rangle$ about \SI{e13}{m^{-2}} which agrees with the typical experimental observation \citep{2013_AM_Jacome, 2000_Superalloy_Miura}. The damage $D$ and plastic shear $\eta^1$ vanish everywhere, as initially there is no damage and plastic activity. The resolved shear stress $\tau^1$ ranges from about \SI{-30}{MPa} in the vertical $\g$ channels to \SI{270}{MPa} in the horizontal $\g$ channels. Note that such a distribution of resolved shear stress contains the contribution of $\ggp$ coherent misfit and external loading. The line tension stress $\tau^1_{\rm l}$ only exists in the $\g$ channels, as the line tension stress mimics the line tension when dislocations are confined in the $\g$ channels. The fraction stress $\tau^1_{\rm f}$ only exists in the $\g$ channels as well and has a uniform value around \SI{16}{MPa}, because we prescribe uniform initial dislocation density in the $\g$ channels. In contrast, the APB stress $\tau^1_{\rm APB}$, which inhibits dislocations from cutting into the $\gp$ precipitates, possesses the $\gp$ precipitates and has a value of \SI{400}{MPa}. Experimental observation shows that dislocation motion mainly takes places in the horizontal $\g$ channels \citep{2000_AM_Link}. This can be expected and explained by the distribution of effective stress $\tau^1_{\rm eff}$ which has values over \SI{150}{MPa} in the horizontal $\g$ channels while vanishes in the vertical $\g$ channels. Besides, experiments show two features during the early creep stage: 1) most dislocations are mobile and multiply in the $\g$ channels, 2) mobile dislocations are blocked by $\ggp$ interfaces and turned into immobile. The former feature is captured in the simulation that initially the multiplication term $\xi^k C_1 (1-\rho_{\rm m}/\rho_{\rm ref})/(d_\g)^4$ dominates over the rest terms of mobile dislocations. The latter feature is captured in the simulation that initially the transfer of mobile dislocations to immobile dislocations due to $\ggp$ interfaces (i.e. $C_3 \rho^1_{\rm m} / (b^1 d_\g)$) dominates over that due to mobile-immobile interactions (i.e. $C_2 \rho^1_{\rm m} \rho^1_{\rm im}$). Note that for the latter feature, experiments show that some of the mobile dislocations may bypass around the $\gp$ precipitates and leave a loop surrounding the $\gp$ precipitates, but the back resistance due to resident loops increases as the resident loops reside more and more, making the dislocation multiplication and bypass increasingly difficult. This process, the resistance in the $\g$ channels increases as the dislocation density increases, is actually captured by the line tension stress term with a prefactor $\rho/\rho_{\rm ref}$. 
  
\begin{figure}[htp] \centering
	\includegraphics[width=\columnwidth]{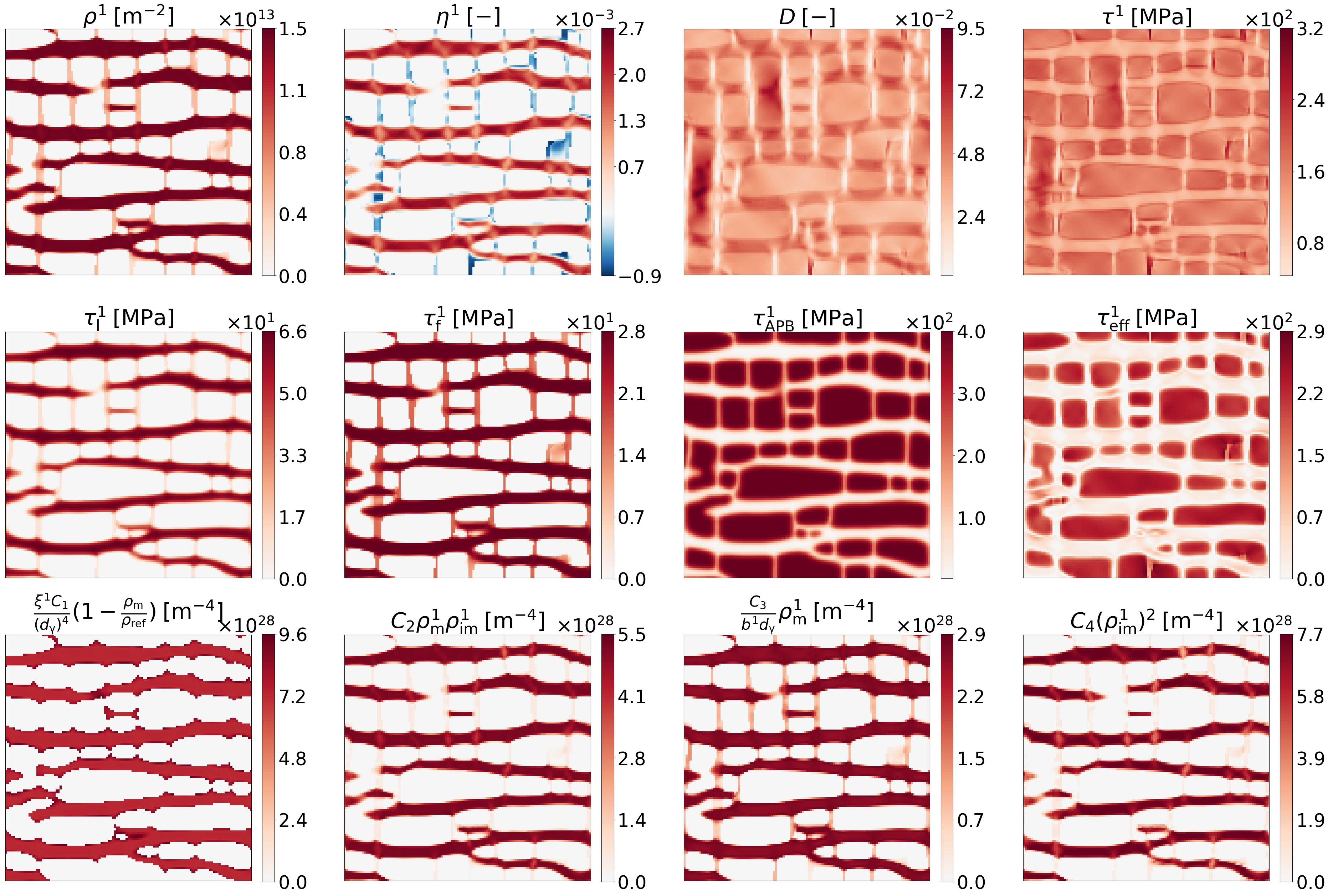}
	\caption{\label{fig: 350MPa_t=2point8h} Dislocation of slip system one and damage associated fields at $\langle \epsilon^{\rm pl}_{22} \rangle=\SI{0.17}{\%}$ and $t=\SI{2.8}{h}$ under \SI{350}{MPa}.}
\end{figure}

After creep of \SI{2.8}{h}, dislocation of slip system one and damage associated fields are shown in \figref{fig: 350MPa_t=2point8h}. It is at early creep stage and the corresponding macroscopic plastic strain along loading direction $\langle \epsilon^{\rm pl}_{22} \rangle$ is \SI{0.17}{\%}. The dislocation density in the horizontal $\g$ channels increases to \SI{1.5e13}{m^{-2}}, while that in the vertical $\g$ channels remains the same. Massive plastic shear can be observed in the horizontal $\g$ channels. Thanks to the treatment explained \figref{fig:swallow-gap}, the "swallow-gap" problem is solved and the dislocation density distribution now agree with experimental observation well. This can be evidently seen in the vertical $\g$ channels where negative plastic shear (i.e. $\eta^1$) is obtained. According to the initial effective stress in \figref{fig: 350MPa_t=0h}, there should be no plastic activity in the vertical $\g$ channels. However, as the $\gp$ precipitates raft to the horizontal direction, some dislocations initially in the vertical $\g$ channels are swallowed by the vertical $\ggp$ interfaces. Then, the APB stress pushes the swallowed dislocation out of the $\gp$ precipitates, leaving negative plastic shear near the vertical $\ggp$ interfaces. Such a strong inhomogeneous distribution of plastic activity dramatically alter the elastic energy distribution, which bring a significant to the rafting of phase microstructures. This can be evidently seen from the damage (i.e. $D$) field. According to the formula of damage, it has strong correlation with the elastic energy distribution. As a consequence of the inhomogeneous dislocation motion, the elastic energy in the horizontal $\g$ channels prevails over that in the vertical $\g$ channels. The $\ggp$ microstructures tend to widen the horizontal $\g$ channels to release the intensive elastic energy and hence cause rafting of $\gp$ precipitates in the horizontal direction. The stress terms (i.e. $\tau^1$, $\tau^1_{\rm l}$, $\tau^1_{\rm f}$, $\tau^1_{\rm APB}$ and $\tau^1_{\rm eff}$) are obviously changed in comparison with their initial fields, due to the evolution of $\ggp$ phases and dislocations. A notable feature is that the effective stress in the horizontal $\g$ channels is reduced compare with the initial state, because dislocation multiplication increases the line tension stress and friction stress. The transfer of mobile to immobile dislocations is also enhanced as the dislocation density increases. Now the transfer terms (i.e. $C_2 \rho^1_{\rm m} \rho^1_{\rm im}$ and $C_3 \rho^1_{\rm m} / (b^1 d_\g)$) of mobile to immobile dislocations increases to the same order of magnitude as the multiplication term (i.e.  $\xi^k C_1 (1-\rho_{\rm m}/\rho_{\rm ref})/(d_\g)^4$), but still smaller than the multiplication term, indicating that dislocation multiplication still goes on with a decreasing multiplication rate. The annihilation of forest immobile dislocations (i.e. $C_4 (\rho^1_{\rm im})^2$) is also enhanced as a result of increment of immobile dislocation density.           

\begin{figure}[htp] \centering
	\includegraphics[width=\columnwidth]{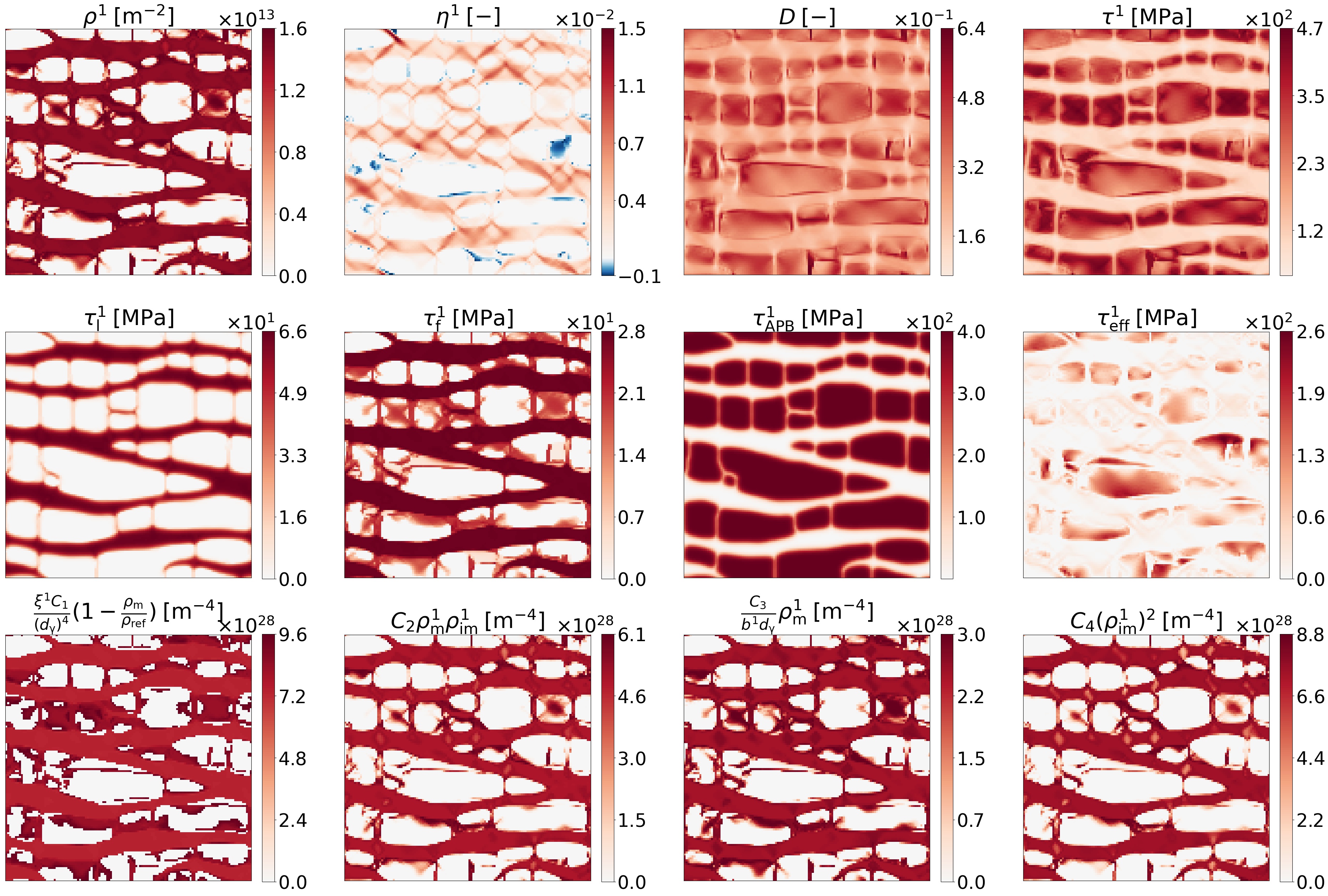}
	\caption{\label{fig: 350MPa_t=20point8h} Dislocation of slip system one and damage associated fields at $\langle \epsilon^{\rm pl}_{22} \rangle=\SI{1.16}{\%}$ and $t=\SI{20.8}{h}$ under \SI{350}{MPa}.}
\end{figure}

After creep of \SI{20.8}{h}, dislocation of slip system one and damage associated fields are shown in \figref{fig: 350MPa_t=20point8h}. It is at the transition between secondary and tertiary creep stages (or the transition between linear and exponential creep) according to the creep curve (see \figref{fig: macroscopic_curve}). The corresponding macroscopic plastic strain along loading direction $\langle \epsilon^{\rm pl}_{22} \rangle$ is \SI{1.16}{\%}. A notable feature is that a few dislocations start cutting into the $\gp$ precipitates. The reason is that now the damage has the maximum value around \SI{0.6}{} inside $\gp$ precipitates, leading to the maximum resolved shear stress around \SI{470}{MPa} which is big enough to overcome the APB stress. The summations of the terms with $C_1 \sim C_3$ and $C_2 \sim C_4$ almost reach balance in the $\g$ channels. This is why the increase rates of dislocation density and plastic shear are low and reach a minimum during the secondary creep.     

\begin{figure}[htp] \centering
	\includegraphics[width=\columnwidth]{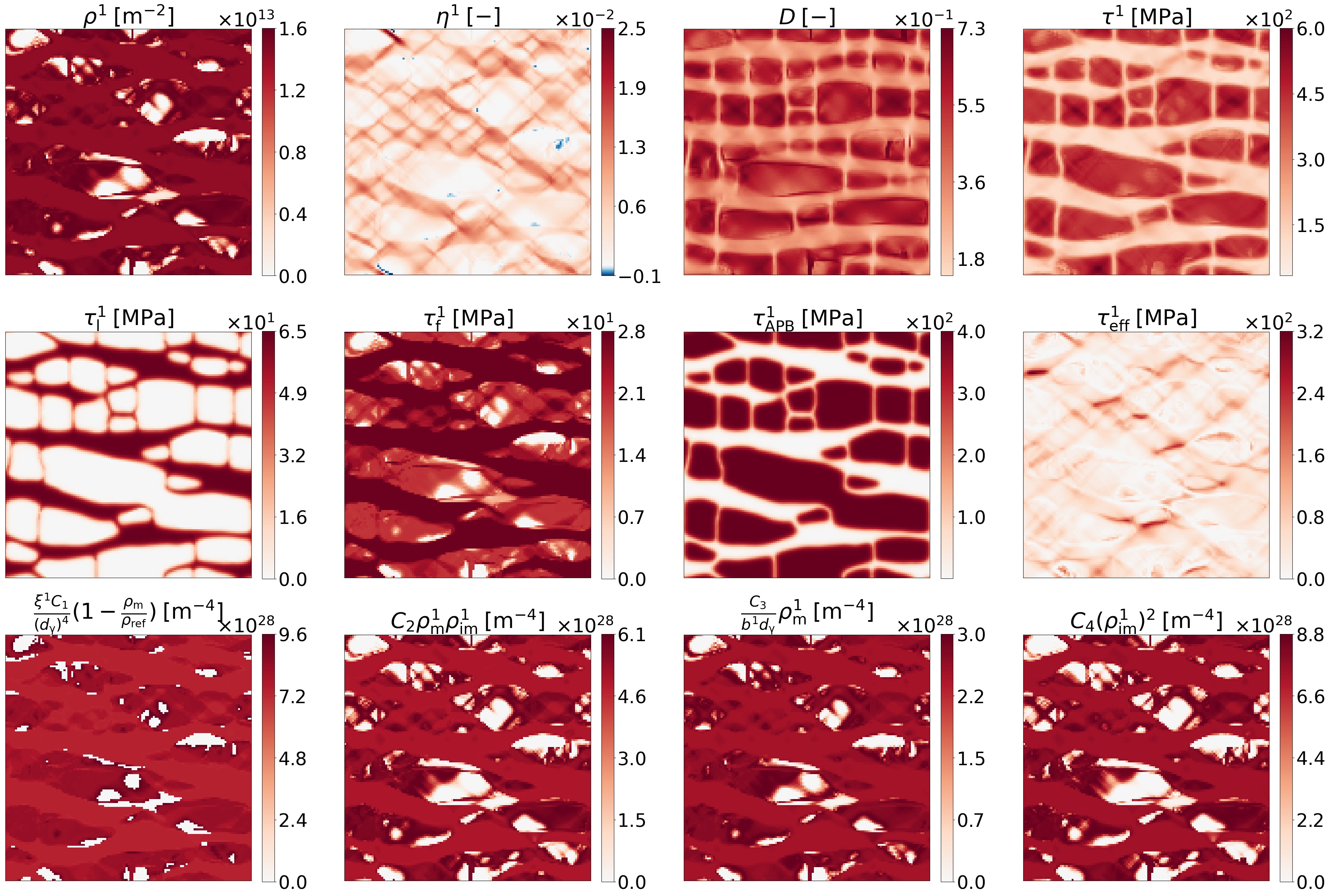}
	\caption{\label{fig: 350MPa_t=30point6h} Dislocation of slip system one and damage associated fields at $\langle \epsilon^{\rm pl}_{22} \rangle=\SI{2.31}{\%}$ and $t=\SI{30.6}{h}$ under \SI{350}{MPa}.}
\end{figure}

After the creep of \SI{30.6}{h}, dislocation of slip system one and damage associated fields are shown in \figref{fig: 350MPa_t=30point6h}. It is obviously at the tertiary creep stage. The corresponding macroscopic plastic strain along loading direction $\langle \epsilon^{\rm pl}_{22} \rangle$ is \SI{2.31}{\%}. Now the resolved shear stress is big enough to over come the APB stress inside most of the $\gp$ precipitates. Because of that, massive dislocations cut into the $\gp$ precipitates can be observed from the dislocation density field. Another interesting feature along with the massive dislocation motion is the formation of slip bands of $30^\circ$ or $60^\circ$ (see the $\eta^1$ field). Such slip bands are typically observed in tensile loading of FCC single crystals when the resolved shear stress is big enough to let dislocation moves in the whole sample area \citep{2005_AM_Westbrooke}. As for the topological inversion, it is not observed at the transition between secondary and tertiary creep stages but obvious during the tertiary creep stage, it seems to suggest the dislocation shearing $\gp$ precipitates is the key factor triggering the topological inversion. When dislocations cut into the $\gp$ precipitates, the elastic energy distribution is changed and such a change affects the evolution the $\ggp$. Anyway, at this point the material has suffered severe degradation and is not qualified for service any more.        

\begin{figure}[htp] \centering
	\includegraphics[width=\columnwidth]{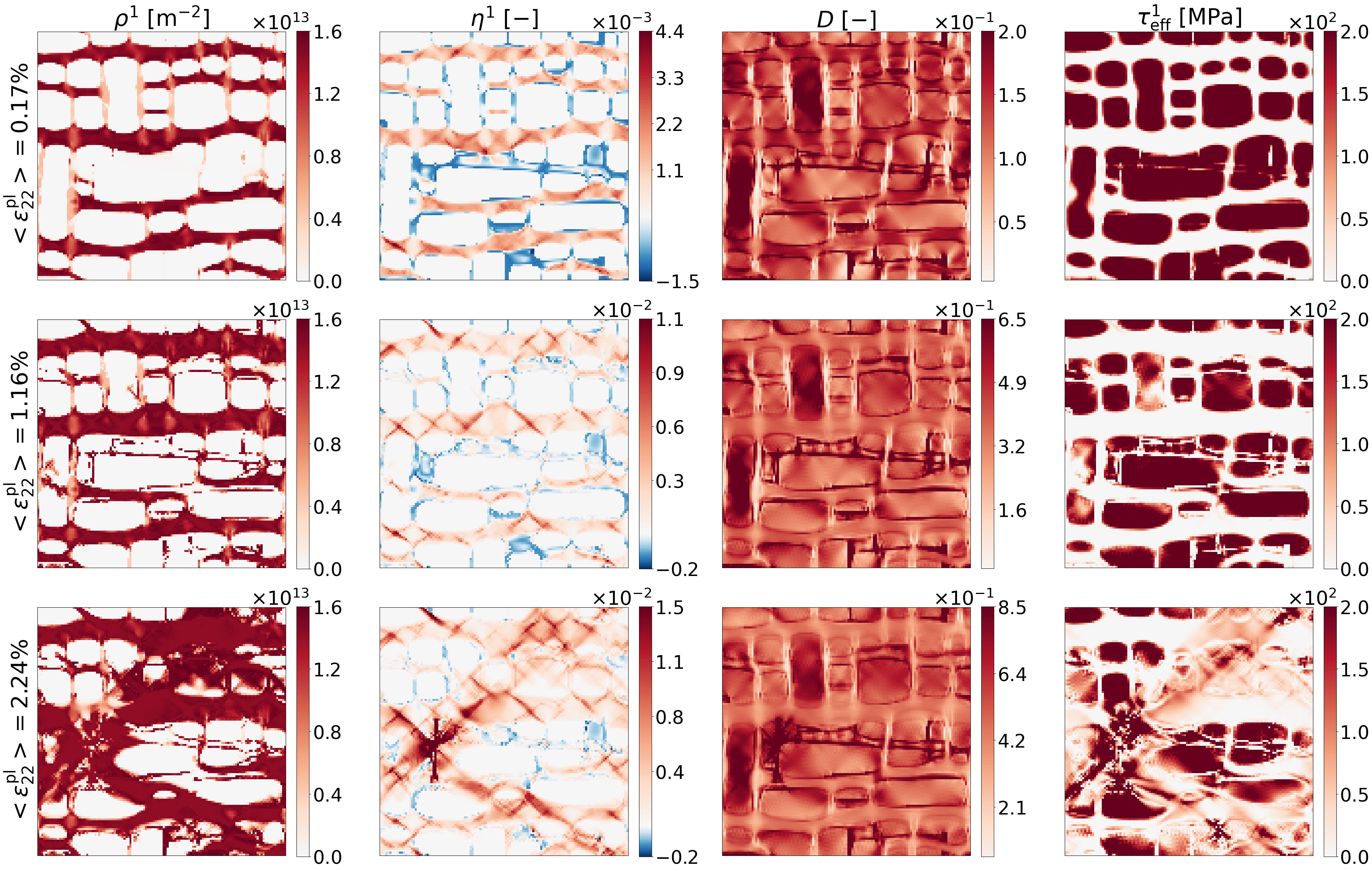}
	\caption{\label{fig: 200MPa_SP1} Dislocation of slip system one and damage associated fields under \SI{200}{MPa}.}
\end{figure}

The dislocation of slip system one and damage associated fields under \SI{200}{MPa} loading are shown in \figref{fig: 200MPa_SP1}. To compare with those under \SI{350}{MPa}, we plot the fields at the similar macroscopic plastic strain (i.e. $\langle \epsilon^{\rm pl}_{22} \rangle=\SI{0.17}{\%}, \langle \epsilon^{\rm pl}_{22} \rangle=\SI{1.16}{\%}$ and $\langle \epsilon^{\rm pl}_{22} \rangle=\SI{2.24}{\%}$). As can be seen, the evolutions of dislocation in slip system one and damage associated fields under \SI{200}{MPa} loading are similar with those under \SI{350}{MPa} loading. It means that as long as it is at the high temperature and low stress loading regime, the microstructural mechanisms are the same under different external stresses. The microstructural images are always similar at the similar macroscopic plastic strain, despite the time reaching the macroscopic plastic strain dramatically increases as the external stress decreases. It can be reasonably deduced from the present results that at high stress loading regime (the external loading is bigger than about \SI{500}{MPa}), the microstructural mechanisms are different from those at low stress loading regimes. Because dislocations can cut into the $\gp$ precipitates and the beginning of deformation. Such a distinction of low stress regime and high stress regime also agree with experiments \citep{2003_JP_Miura, 2007_AM_Rae, 1999_AM_Matan}.    

\begin{figure}[htp] \centering
	\includegraphics[width=\columnwidth]{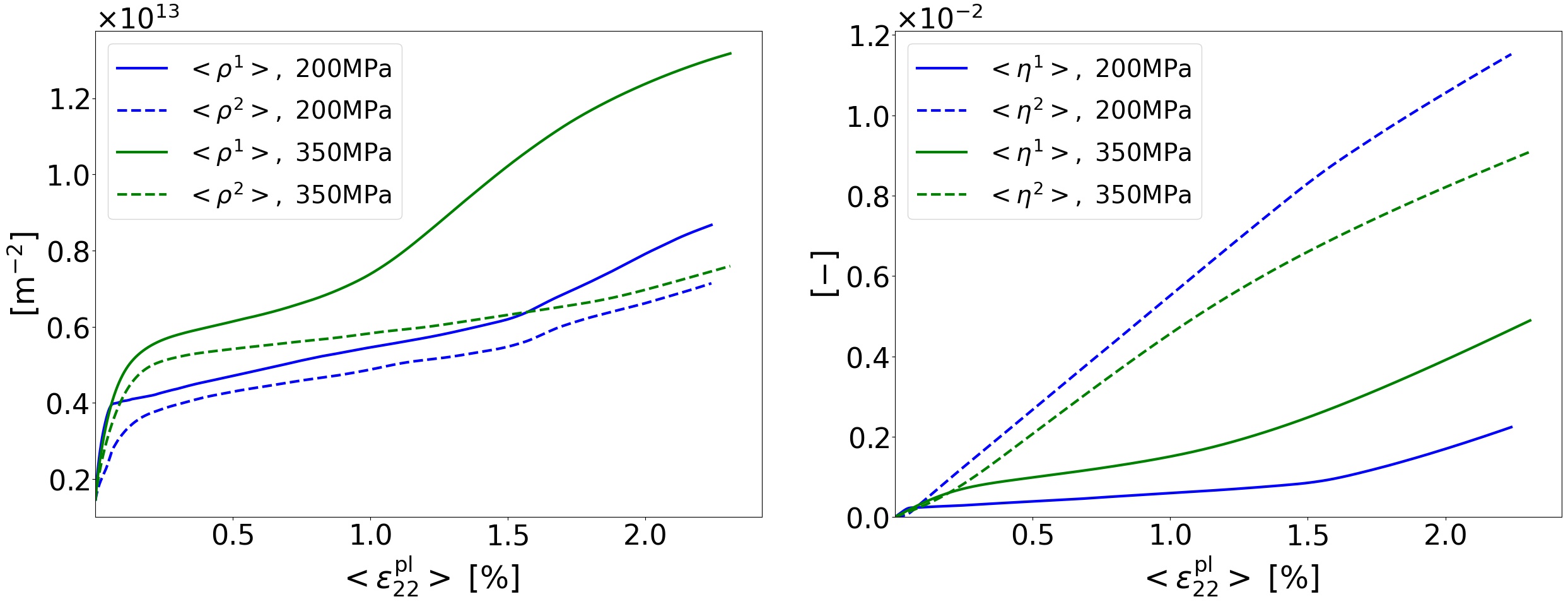}
	\caption{\label{fig: 200_and_350MPa_SP1_and_SP2_compare} Comparison between results of slip system one and two: left, dislocation density; right, plastic slip.}
\end{figure}

\begin{figure}[htp] \centering
	\includegraphics[width=\columnwidth]{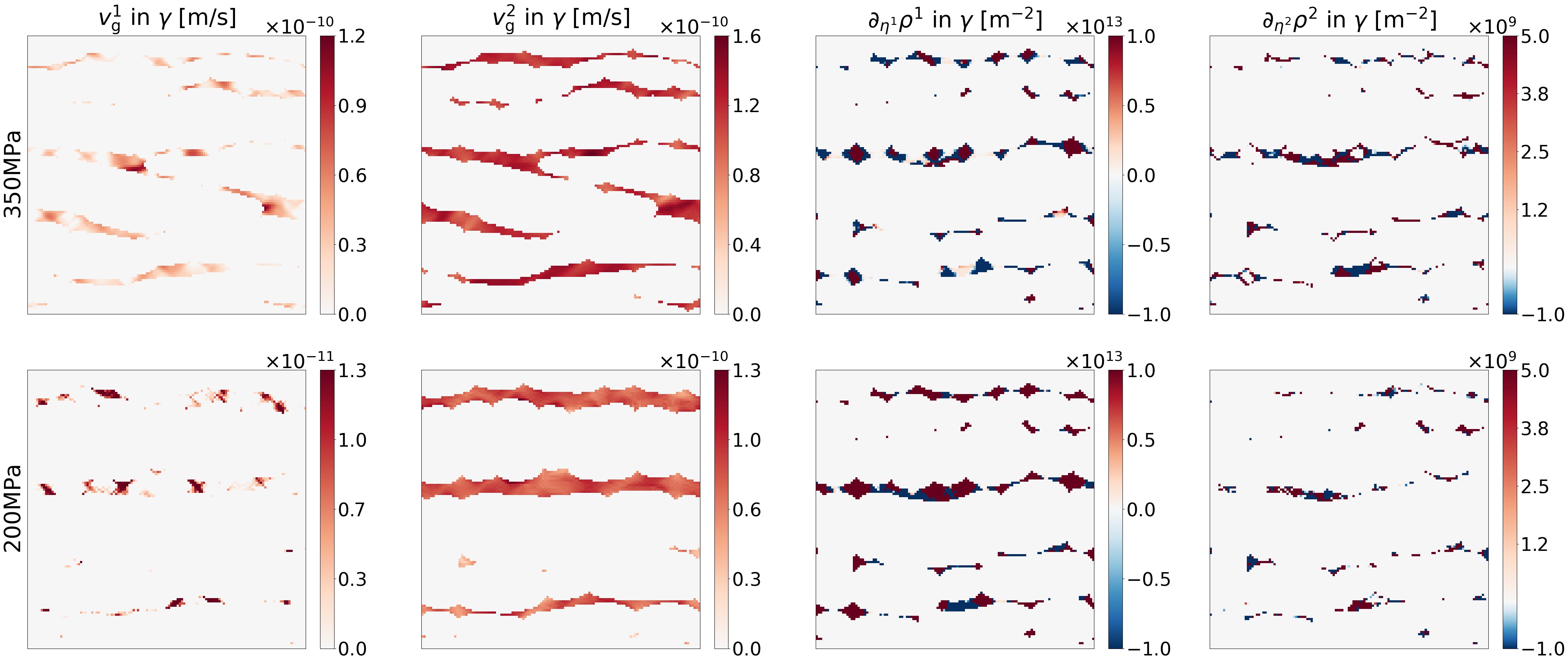}
	\caption{\label{fig: 200_and_350MPa_vg} Comparison of fields of slip system one and two under \SI{200}{MPa} and \SI{350}{MPa} at $\langle \epsilon^{\rm pl}_{22} \rangle=\SI{1.16}{\%}$.}
\end{figure}

The dislocation evolutions of odd slip systems are quantitatively very similar with each other, so do those of the even slip systems. The evolution mechanisms of odd and even slip systems are qualitatively the same, but have some distinctions quantitatively, as shown in \figref{fig: 200_and_350MPa_SP1_and_SP2_compare}. The macroscopic value of a quantity can be obtained by doing spatial average of that quantity by $\langle X \rangle = \int X \dV / V$. It is not a surprise that there are more dislocations under \SI{350}{MPa} than under \SI{200}{MPa}. However, it is interesting that there are less dislocations but more plastic shear in the even slip systems than in the odd slip slip systems, regardless of the external stress. This can be explained from the fields plotted in \figref{fig: 200_and_350MPa_vg}. The rafted phase microstructures result in bigger resolved shear stress in the slip system two than in the slip slip one, which consequentially leads to bigger dislocation evolution velocity and plastic shear (see \eqref{eq: Orowan shear rate final}) in the slip system two. At first glance, it seems that bigger dislocation evolution velocity leads to bigger dislocation density in the slip system two as well according to \eqref{eq: evolution rate of mobile dislocation density extended} and \eqref{eq: evolution rate of immobile dislocation density extended}. Nevertheless, the evolution of dislocation density also depends on the $\partial_\eta^k \rho^k$. It can be seen that $\partial_{\eta^1} \rho^1$ is much bigger than $\partial_{\eta^2} \rho^2$ regardless of the external stress ($\langle \partial_{\eta^1} \rho^1 \rangle = $ \SI{7.6e14}{m^{-2}} and $\langle \partial_{\eta^2} \rho^2 \rangle = $ \SI{4.8e14}{m^{-2}} for \SI{350}{MPa}, $\langle \partial_{\eta^1} \rho^1 \rangle =$ \SI{3.4e16}{m^{-2}}  and $\langle \partial_{\eta^2} \rho^2 \rangle = $ \SI{3.9e15}{m^{-2}} for \SI{200}{MPa}), leading to bigger dislocation density in the slip system one than in the even slip system two.         
 
\subsection{Macroscopic properties}
\begin{figure}[htp] \centering
	\includegraphics[width=\columnwidth]{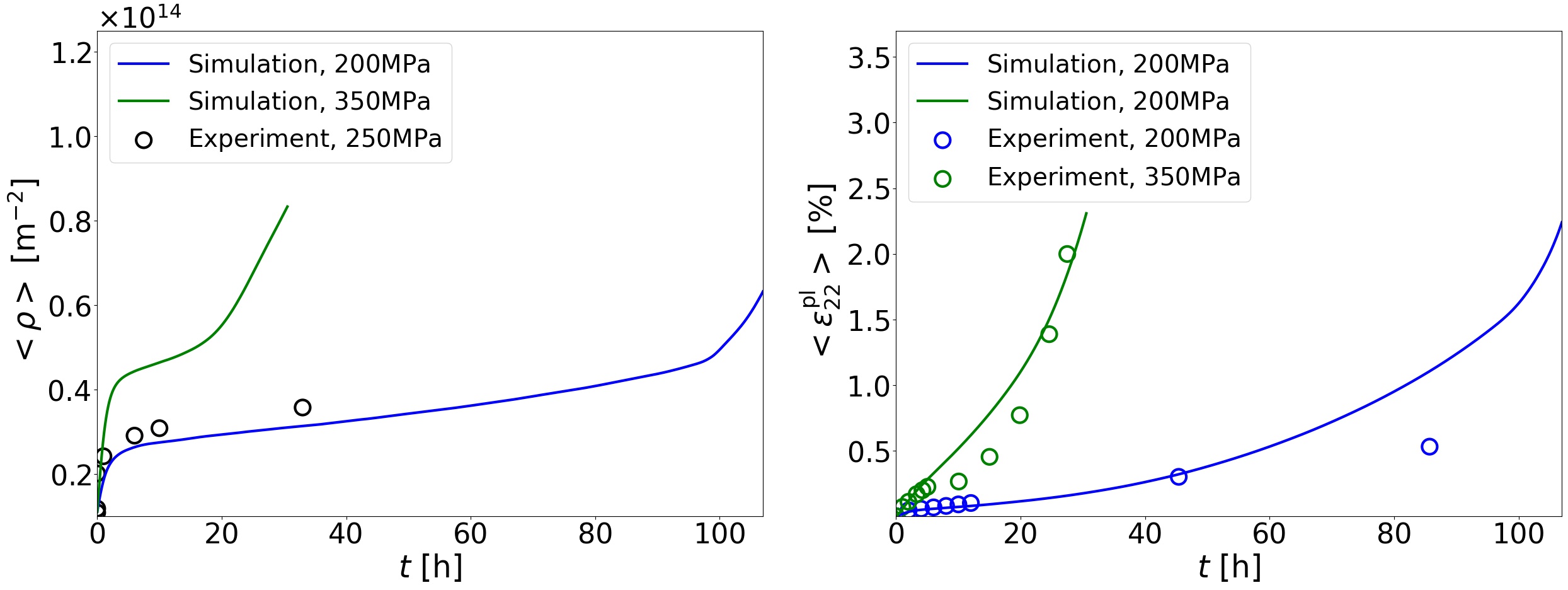}
	\caption{\label{fig: macroscopic_curve} Comparison between simulated and experimental results: left, dislocation density; right, plastic strain in loading direction. Experimental data are from \citep{2014_Conference_Kondo, 2000_AM_Link}.}
\end{figure}

The simulated temporal evolutions of macroscopic creep properties are shown in \figref{fig: macroscopic_curve} and compared with experimental data. As can be seen, the simulated shape and value of the curves principally agree with the experimental results \citep{2014_Conference_Kondo, 2000_AM_Link}. The increments of dislocations and creep strain during early creep stage are due to the multiplication of dislocation in the $\g$ channels. The increments of dislocations and creep strain during tertiary creep stage are due to the dislocation cutting into $\gp$ precipitates. The minimum creep rates under \SI{350}{MPa} and \SI{350}{MPa} are \SI{9e-9}{s^{-1}} and \SI{1.3e-7}{s^{-1}}, respectively, which are very close to experimental data under similar loading conditions \citep{2000_Superalloy_Miura}. 

Although we present the creep simulation of single crystal superalloys here, the coupled framework can be applied to polycrystals with some modifications. Firstly, a set of additional order parameters, for instance $\psi_i$ where $i$ indicates different grains, should be added to the phase-field model. Then, the $d_\g$ might be replaced by the grain size to recover the Hall-Petch hardening due to grain boundaries. The APB stress might be replaced by the stress needed to nucleate dislocations from grain boundary. Since we use a relatively simple description for the damage, it can be improved by a more physical way (see \citep{2015_CM_Ambati} for example). We may improve the current model and extend its application following these ideas in the future.                  

\section{Conclusions}
\label{sec:Conclusion}
A phase-field, dislocation density based plasticity and damage coupled model is developed and applied to the high temperature creep of single crystal superalloys under \SI{200}{MPa} and \SI{350}{MPa}. The following conclusions can be drawn from the present work: 
\begin{enumerate}[(1)]
	\item The simulated spatial-temporal evolutions of phase microstructures during the whole creep stage agree with experiments well. Some key features like rafting and topological inversion of phase microstructures can be captured. The strong inhomogeneous plastic activity (a lot in the horizontal but a little in the vertical $\g$ channels) is the main reason causing rafting, apart from the inhomogeneous elasticity. The dislocation cutting into $\gp$ precipitates is likely to be the main reason triggering topological inversion of phase microstructures.       
	\item The simulated spatial-temporal evolutions of dislocations during the whole creep stage principally agree with experiments. Some representative features like dislocation multiplication in the $\g$ channels, mobile-immobile transition due to block of $\ggp$ interfaces and cutting into $\gp$ precipitates are observed in the simulations. Especially, the "swallow-gap" problem is solved in the present model. The even slip systems produce less dislocations but more plastic slip than the odd slip systems, if the slip systems are defined as the present work.        
	\item The simulated macroscopic properties including the minimum creep rate, dislocation density and creep curves agree with experiments. The increments of dislocations and creep strain during early creep stage are due to the multiplication of dislocation in the $\g$ channels. The increments of dislocations and creep strain during tertiary creep stage are due to the dislocation cutting into $\gp$ precipitates. 
	\item Since the simulated microscopic and macroscopic evolutions all principally agree with experiments, it is fair to state that the kinematics and dynamics of present model are principally reasonable. The present model is applicable for understanding and predicting the creep microstructures and properties of single crystal superalloys.

\end{enumerate}
\section*{Acknowledgment}
The research is financially supported by National Natural Science Foundation of China (12002275) and Natural Science Foundation of Shaanxi Province (2020JQ-125).

\bibliography{Manuscript}

\end{document}